\documentclass[conference]{IEEEtran}
\IEEEoverridecommandlockouts
\usepackage{cite}
\usepackage{amsmath,amssymb,amsfonts}
\usepackage{algorithmic}
\usepackage{graphicx}
\usepackage[scaled=0.92]{inconsolata}
\usepackage{textcomp}
\usepackage{xcolor}
\usepackage{verbatim}
\usepackage{cleveref}
\usepackage{pgfplots}
\pgfplotsset{compat=newest}

\def\BibTeX{{\rm B\kern-.05em{\sc i\kern-.025em b}\kern-.08em
   T\kern-.1667em\lower.7ex\hbox{E}\kern-.125emX}}

\usepackage{listings}
\definecolor{dkgreen}{rgb}{0,0.6,0}
\definecolor{gray}{rgb}{0.5,0.5,0.5}
\definecolor{mauve}{rgb}{0.58,0,0.82}

\def \APP {CapExec}
\newcommand{\name}[1]{\lstinline{#1}}

\begin{document}

\title{\APP{}: Towards Transparently-Sandboxed Services (Extended Version)}

\author{
  \IEEEauthorblockN{
    Mahya Soleimani Jadidi\IEEEauthorrefmark{1},
    Mariusz Zaborski\IEEEauthorrefmark{2},
    Brian Kidney\IEEEauthorrefmark{1},
    Jonathan Anderson\IEEEauthorrefmark{1}
  }
  \\
  \IEEEauthorblockA{
    \IEEEauthorrefmark{1}
    \textit{Department of Electrical and Computer Engineering} \\
    \textit{Memorial University of Newfoundland}\\
    \{msoleimanija,brian.kidney,jonathan.anderson\}@mun.ca
  }
  \\
  \IEEEauthorblockA{
    \IEEEauthorrefmark{2}
    \textit{Research and Development} \\
    \textit{Fudo Security Inc.}\\
    \{oshogbo@FreeBSD.org\}
  }
}

\maketitle
\IEEEpeerreviewmaketitle

\begin{abstract}

Network services are among the riskiest programs executed by production systems.
Such services execute large quantities of complex code and process data from arbitrary --- and untrusted --- network sources, often with high levels of system privilege.
It is desirable to confine system services to a least-privileged environment so that the potential damage from a malicious attacker can be limited, but existing mechanisms for \textit{sandboxing} services require invasive and system-specific code changes and are insufficient to confine broad classes of network services.

Rather than sandboxing one service at a time, we propose that the best place to add sandboxing to network services is in the \textit{service manager} that starts those services.
As a first step towards this vision, we propose \APP{}, a process supervisor that can execute a single service within a sandbox based on a service declaration file in which, required resources whose limited access to are supported by Caper services, are specified. 
Using the Capsicum compartmentalization framework and its Casper service framework, \APP{} provides robust application sandboxing without requiring any modifications to the application itself.
We believe that this is a first step towards ubiquitous sandboxing of network services without the costs of virtualization.
\end{abstract}


\section{Introduction}
\label{intro}
Network services and applications have always been attractive targets for remote attackers.
Network services typically incorporate complex protocol parsing code, often written in low-level languages, that is exposed to arbitrary content from the network.
Since these services commonly execute with system privilege, they are at high risk for remote exploitation as a gateway to other system resources.
Because of the risk and the consequence of potential compromises, there is a need to confine network applications and limit the damage that can be inflicted by a successful attack.

One broad class of techniques that seem applicable to the problem of securing network services is \textit{sandboxing}: restricting software's access to system resources such that the application has the least privilege required to fulfill its function.
However, applying such limitations is challenging.
Many sandboxing frameworks require invasive code modifications, some of which require a great deal of security expertise to apply correctly. 
Incorrectly applied mechanisms may lead to a false sense of security without additional effectual protection.
What is needed, in addition to the development of effective sandboxing techniques, is the development of tools to apply those techniques.

Sandboxing is applicable through different techniques at various levels within operating systems such as system call \name{chroot(2)}\cite{chroot}, sandboxing features in service managers such as \name{systemd}\cite{systemd}, or application containers such as \name{jail(8)}\cite{jail} or \name{docker}\cite{docker}. We believe that a great deal of benefit can be derived from the application of sandboxing at the level of system services without requiring invasive modifications to the application code of every network service. Hence, we see \textit{service managers} as the key to securing systems with network-facing services. Securing systems at this level, eliminates the need for securing every service separately.

In this paper, we have combined process supervision and FreeBSD's capability-oriented compartmentalization framework\cite{freebsd-di}, Capsicum~\cite{capsicum}, to introduce \APP{} as a sandboxing process supervisor (\cref{sec:the-app}), which runs applications in a restricted \textit{capability} mode.

Capsicum provides a fine-grained sandboxing mechanism, giving developers a significant degree of control within an application.
However, some developers still avoid the use of this technique, or even other sandboxing approaches, for the following reasons:
\begin{itemize}
    \item Difficulties of preserving the functionality of the program while enhancing its security
    \item Requirement for source code modifications and the additional associated testing
    \item Verifying the correctness and compatibility between limits defined in the sandbox
\end{itemize}

These issues motivated us to facilitate the use of Capsicum by employing the same ideas in our supervisor program.

\APP{} uses Capsicum to restrict the privileges of services, limiting access to resources from compromised applications; it uses the Capsicum-based Casper service framework~\cite{casper-daemon, libcasper} to provide access to required resources through capability channels defined by Casper.
To use  \APP{}, services' required run-time resources should be described in service declaration files, as an initial step to unify security and functionality descriptions.
When a service's run-time requirements can be described in terms of Casper services, {\APP} provides sandboxing without any modifications to source code. 
The mechanism and its various evaluation results are described in  \cref{sec:the-app}  and \cref{evaluation} respectively.
This project is a significant advance beyond the current state of the art (\cref{sec:related-work}).

%This sandboxing supervisor is the start of a research effort to bring generally-applicable sandboxing to the highest-risk applications in an approach that is \textit{transparent}, i.e., without requiring any modifications to application code (\cref{sec:future-work}).

\section{{\APP}: A Sandboxing Service Supervisor}
\label{sec:the-app}

We have designed and developed {\APP} to be a security-focused service supervisor that executes a service in a sandbox transparently, which means no modification to the service's source code is required.

{\APP} creates sandboxes relying on Capsicum\cite{capsicum} and Casper services.
Capsicum is a sandboxing framework developed for FreeBSD. The fundamental idea for the framework comes from the concept of capabilities in capability-based systems \cite{capability-based-systems}. Capabilities are unforgeable tokens of authority carried by processes to authorize access to the system’s resources. Using Capsicum’s API, once a process enters to the capability mode, all system calls trying to access global namespaces, such as the root filesystem or the PID\footnote{Process ID} namespace, will fail.
Capsicum’s sandboxing has changed the regular flow of the system call mechanism in FreeBSD. In addition to Capsicum’s general API, Casper has extended Capsicum’s features by defining capability channels to allow access to some riskier but widely-used services\cite{casper-daemon}.

Using Capsicum and Casper, {\APP} creates and loads required Casper services to provide access to white-listed run-time resources. These services and their limits are specified in the application's declaration files as the service configuration. {\APP} executes one application at the time in a sandbox made with required Casper's capability channels. 
Hence, limited accesses to the subsetted namespaces are proxied for the application, but any unprivileged behaviour or requests beyond defined limits will fail due to capabilities violations.

\subsection{Service Declaration: Unifying Application Functionality with Security Requirements}
\label{service-declaration}

\APP{} employes \name{libucl}\cite{ucl} to parse the service declaration file conforming to JSON format\cite{json}. The file should describes the service's binary and its \textit{requirements}.
\textit{Requirements} are those system resources that are disallowed in capability mode, but their alternative restricted services are supported by Casper daemon or \name{libcasper} API\cite{libcasper}. 
An example configuration for the \name{traceroute} utility is given in Listing \ref{traceroute-service-declaration}.

\begin{lstlisting}[language=csh, caption=An example of the content of traceroute declaration file, label={traceroute-service-declaration}]
{
        binary: "/usr/sbin/traceroute"
        "system.fileargs" : {
                operations: "OPEN",
                flags: "RDONLY",
                cap_rights: "READ",
                cap_rights: "FCNTL",
                cap_rights: "FSTAT",
                filename: "/etc/protocols",
                filename: "/dev/null"
        }
        "system.dns" : {
                family: AF_INET
        }
        "system.net" : {
                host: "example.com",
                family: AF_INET
        }
        "system.sysctl" : {
                "vm.overcommit": {
                        type: "mib",
                        flag: "CAP_SYSCTL_READ"
                }
        }
}
\end{lstlisting}

\begin{figure}[ht]
\centering
\includegraphics[scale=0.37]{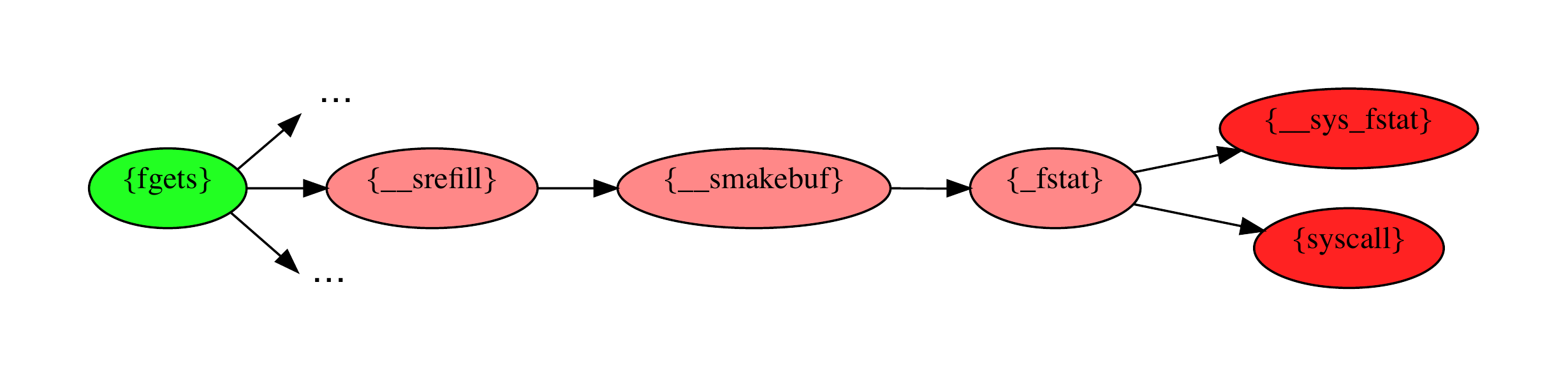}
\caption{Reduced call graph from \name{traceroute(8)} showing how a call to \name{fgets(3)} can result in system calls disallowed in Capsicum.}
\label{fig.graph}
\end{figure}

Writing service declarations requires knowledge of system calls used within the applications code. The problem is complicated by system calls used indirectly by libraries.
Recognizing this issue, we developed \name{CapCheck}, a tool for highlighting library calls that use system calls not allowed in a Capsicum sandbox. As can be seen in figure \ref{fig.graph}, even a call to \name{fgets(1)} can result in unexpected system calls that are disallowed in Capsicum sandboxes.

\name{CapCheck} uses \name{readelf} and \name{ldd} to determine the calls made to external libraries and the libraries that provide them.
It then builds a full call graph for the application and searches it to find paths that result in disallowed system calls.
This information is provided to the end user to develop service declaration files.

\subsection{Sandboxed Execution}
\label{Mahyad-internal-design}

\begin{figure}[ht]
\centering
\includegraphics[scale=0.7]{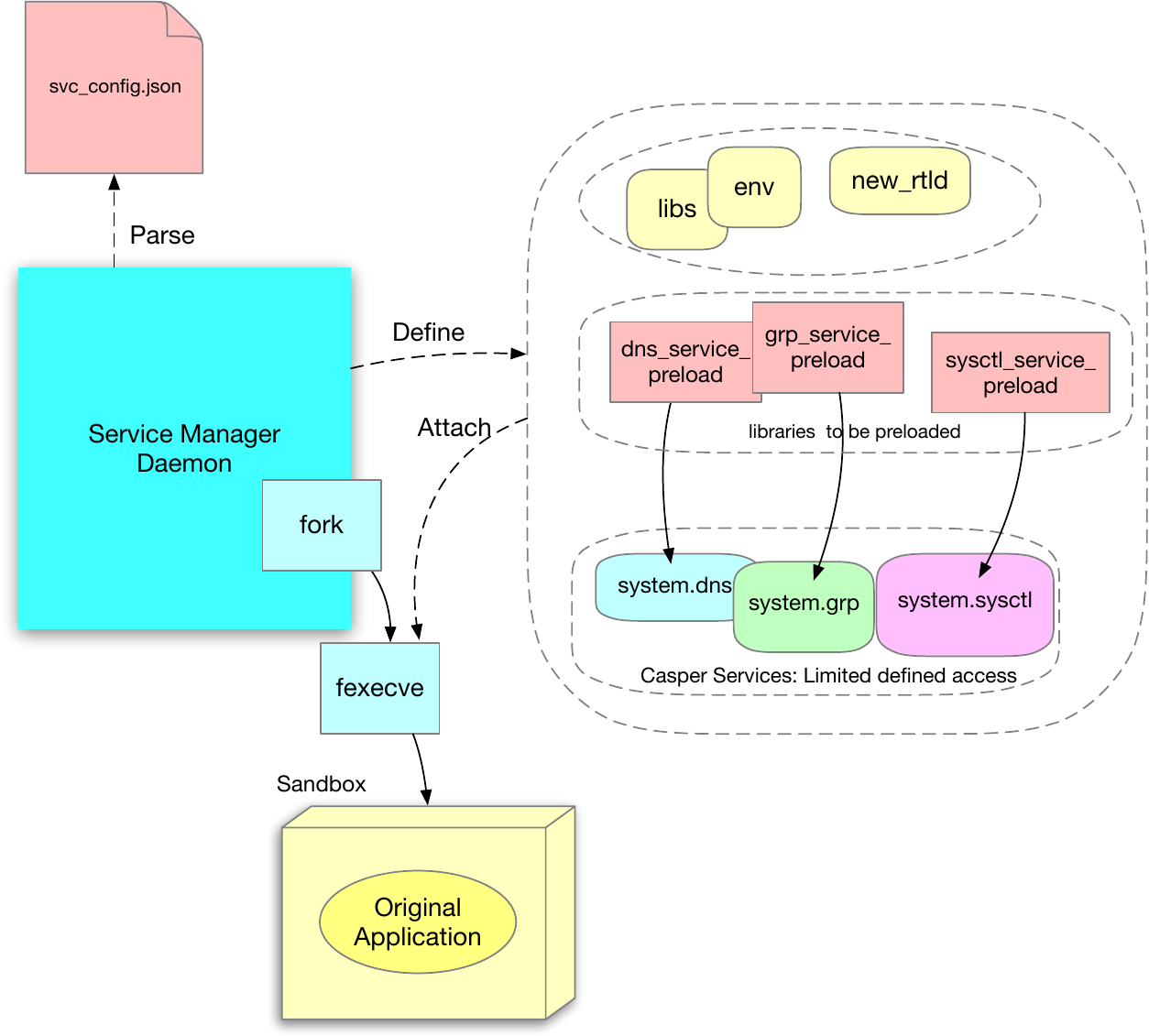}
\caption{{\APP}'s approach to run a service in a sandbox}
\label{mahyad}
\end{figure}

As the first step, {\APP} parses declaration files and creates the corresponding Casper services. In addition to supporting all Casper services, {\APP} needed a capability channel for simple network communications. Therefore, we developed an experimental networking Casper service, \name{system.net}, supporting \name{bind(2)} and \name{connect(2)}. After parsing service declaration files, \APP{} forks and executes the binary sandboxed in a child process, creating a new environment or context for the process, including arguments, libraries, environment variables, the new runtime linker, shared memory mappings and required libraries to make the process sandboxed.
{\APP} replaces disallowed \name{libc} system calls with other functions that redirect requests to the existing Casper services.
As a result, the source code remains unmodified.
Figure \ref{mahyad} shows this mechanism.

\section{ {\APP} in practice}
\label{case-studies}

A useful approach to achieve compartmentalization in large networks is to employ communication and access policies on subnets.
There are always applications such as firewalls and similar tools to enforce policies in a coarse-grained form of actions being held on the edge of the system.
{\APP} is a solution to have finer-grained enforced policies on services rather than users behaviors.
If administrators can define detailed policies for important system services, then we will have network services that might be compromised, but they will not cause information leakages over the network, privilege escalation on servers or other consequence server failures.
{\APP}'s dependency on configuration files causes administrators to be able to run sandboxed network services using simple network management tools such as running startup or shut down scripts on specific domains. Limits can be applied to the network more manageable, centralized, and more verifiable.

{\APP} turns application sandboxing from a complicated and time-consuming procedure into one that is simpler and faster to implement.
To examine this set of tools we sandbox two of well-known existing utilities, \name{cat(1)} and \name{traceroute(8)}, on FreeBSD 12.0.
In this section, we describe \APP{}'s internal mechanism.

\subsection*{What happens to the process?}
\label{capexec-details}

Using {\APP}, the administrator or developer of the system can describe required resources in terms of Casper's services similar to the example that has been shown in the listing \ref{traceroute-service-declaration}.
Suppose that listing \ref{local-system-sample} is a piece of code for our process.

\begin{figure}[ht]
\centering
\includegraphics[scale=0.5]{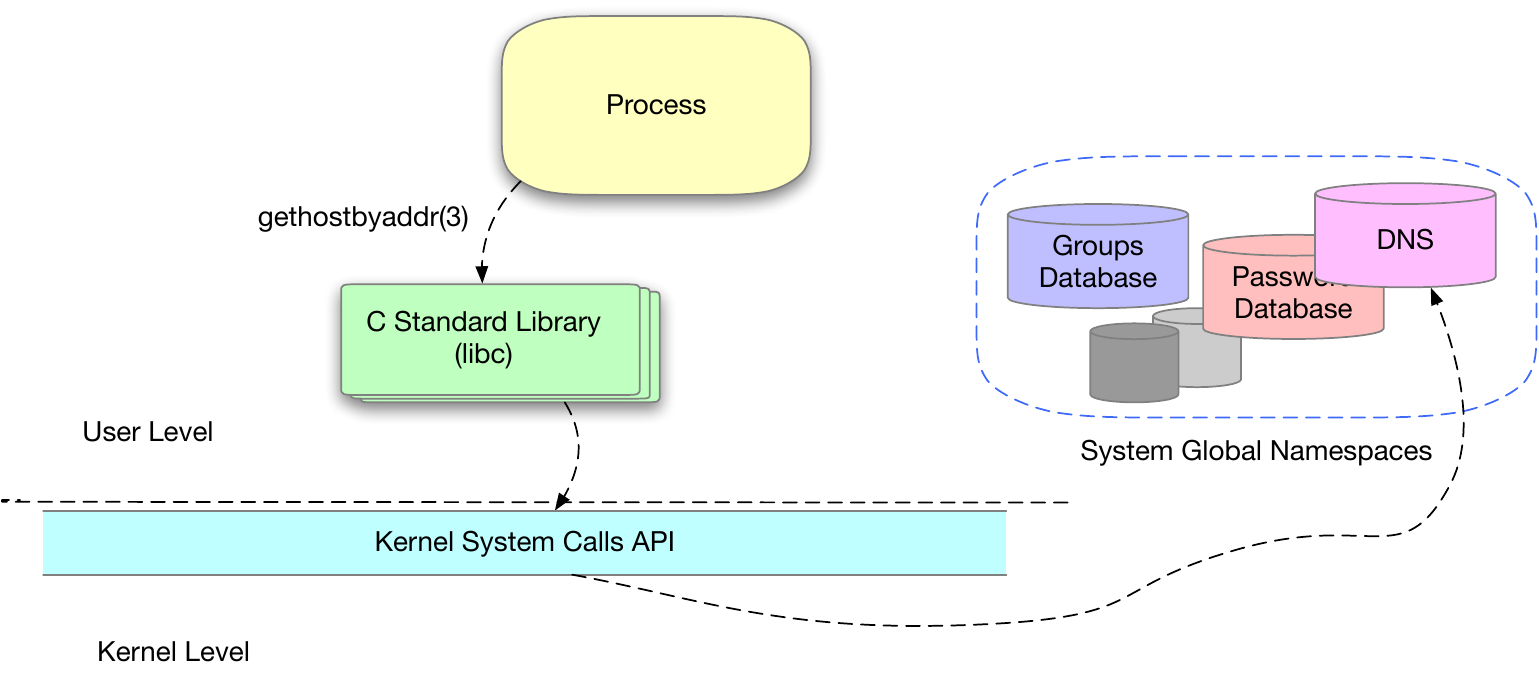}
\caption{The original procedure of calling unsafe functions and accessing to the system global namespaces}
\label{no-sandbox}
\end{figure}

\begin{lstlisting}[language=C++, caption=Software that depends on functionality disallowed by Capsicum, label={local-system-sample}]
void  test_gethostbyname() {
  const char *ipstr = "127.0.0.1";
  struct in_addr ip;
  struct hostent *hp;
  if(!inet_aton(ipstr, &ip))
    fprintf(stdout, "Unable to parse IP address %s.", ipstr);

  hp = gethostbyaddr((const void*)&ip, sizeof(ip), AF_INET);
  CPPUNIT_ASSERT(hp != NULL);
}
\end{lstlisting}
In the absence of {\APP}, the process calls \name{gethostbyaddr(3)} function and the corresponding \name{libc} function will be called as shown in figure \ref{no-sandbox}.

\begin{figure}[ht]
\centering
\includegraphics[scale=0.5]{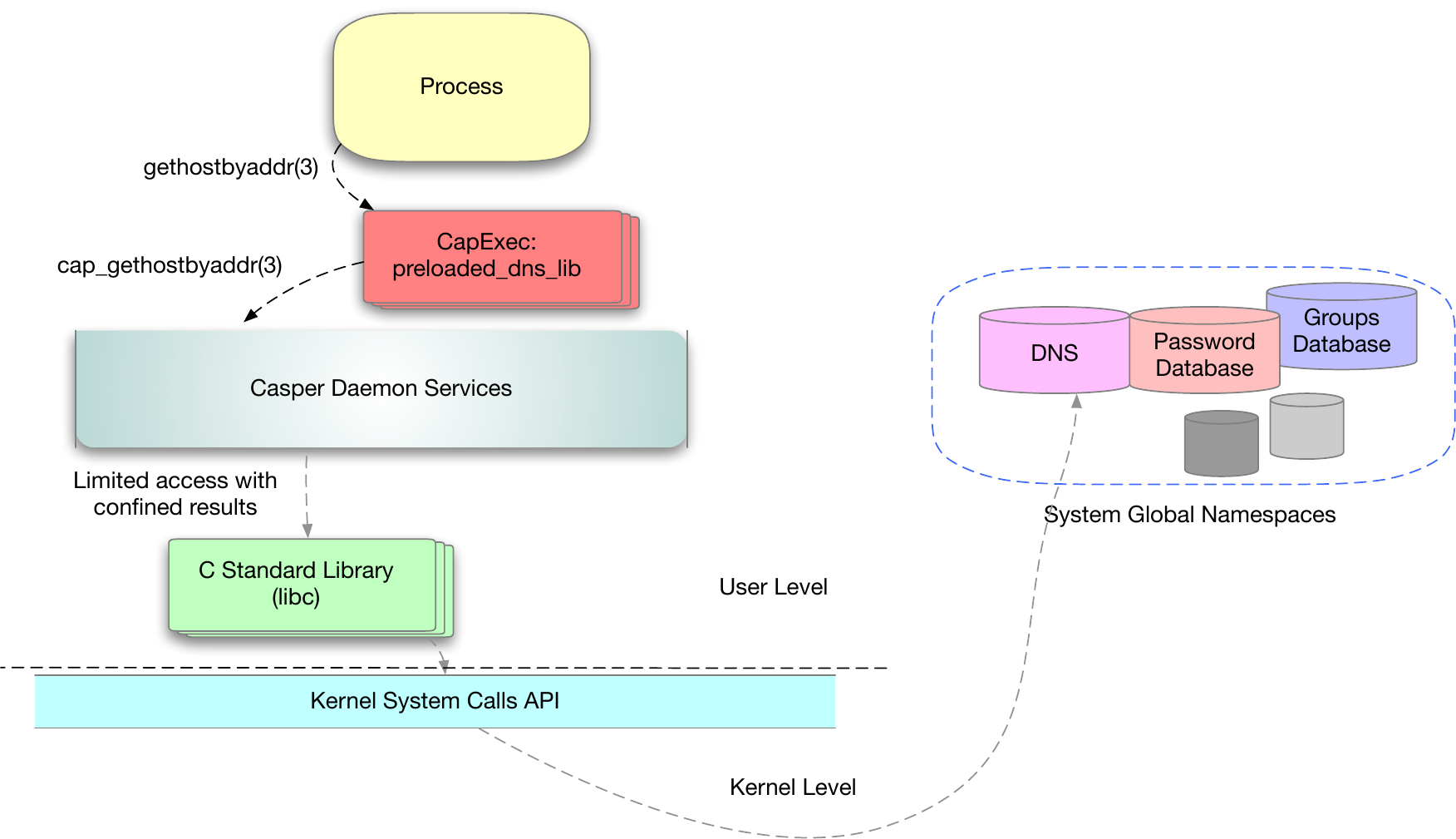}
\caption{The procedure of process execution with {\APP}.
The process disallowed invocations are redirected to predefined capability channel in the corresponding Casper service. Numbers are showing the order of steps.}
\label{fig.process-sandbox}
\end{figure}

When a process enters the capability mode, all system calls which try to access the global namespaces will fail.
So the corresponding system functionality will not be met.
The Casper daemons help solve this problem by creating restricted capability channels to access system resources with limited privileges.
To use Casper services in {\APP}, the service configuration file should contain the following content for our example:

\begin{lstlisting}[language=csh, caption=An example of a service declaration for \name{system.dns}  ]
{
	"system.dns": {
		type: ADDR,
		family: AF_INET
	}
}
\end{lstlisting}

When {\APP} parses the file, it defines a capability channel for \name{system.dns} service with specified limits and sets required libraries to be preloaded later.
Preloaded libraries and their functions retrieve capability channels that are open in the parent process, which is the main thread of {\APP}.
In our example,  a capability channel will be established for the service \name{system.dns}.
Then calling \name{gethostbyaddr(3)} will be redirected to the channel with Casper's equivalent function \name{cap\_gethostbyaddr(3)}.
The procedure is shown in figure \ref{fig.process-sandbox}.

As it is mentioned before, to test our application functionality on elementary utilities, we tried to sandbox \name{cat(1)} and \name{traceroute(8)}.
\name{cat(1)} is a very small utility which needs only one Casper service, \name{system.fileargs}, to open files specified as command line arguments.
Listing \ref{cat.svc} shows \name{cat}'s service declaration file that we used to make it sandboxed.

\begin{lstlisting}[language=csh, caption=An example of a service declaration to sandbox \name{cat(1)} , label=cat.svc ]
{
	binary: "/bin/cat"
	"system.fileargs": {
		operations: "OPEN",
		flags: "O_RDONLY",
		cap_rights: "READ",
		filename:  "test.txt"
	}
}
\end{lstlisting}

\section{Evaluation and Comparisons}
\label{evaluation}

\subsubsection{Runtime Performance}
In this section, we describe evaluation results and our observations about runtime performances, memory usage, correctness, and comparison of running existing applications, \name{cat(1)} and \name{traceroute(8)}, in \APP{}'s sandbox and other sandboxing technologies. 
We have chosen these two utilities as they are both simple small utilities sandboxed before using Capsicum.
Examining these application allows us to have a direct comparison to traditional applications of Capsicum requiring source code modification. 

\begin{figure}[ht]
\begin{tikzpicture}[xscale=0.75, yscale=0.75]
\begin{axis}[
	ymin=0,
	ylabel={Time (s)},
	ymode = log,
	xlabel={Size in MBs},
	xtick={1, 2, 3, 4, 5},
	xticklabels={1, 10, 100, 500, 1000},
	legend entries={cat, cat with CapExec},
        legend pos= south east
]
\addplot+[error bars/.cd,y dir=both,y explicit]
coordinates {
	(1, 0.002)  +- (0, 0)
	(2, 0.01)  +- (0,0)
	(3, 0.105)
	(4, 0.665)      +- (0,0.0)
	(5, 1.34)      +- (0,0.005)
};
\addplot+[error bars/.cd,y dir=both,y explicit]
coordinates {
	(1, 0.01) +- (0, 0)
	(2, 0.02) +- (0, 0)
	(3, 0.12) +- (0, 0)
	(4, 0.68) +- (0, 0.1)
	(5, 1.36) +- (0, 0.2)
};
\end{axis}
\end{tikzpicture}
\caption{Time to open single file with original \name{cat(1)} in comparison with cat running in CapExec's sandbox}
\label{cat-single-file}
\end{figure}
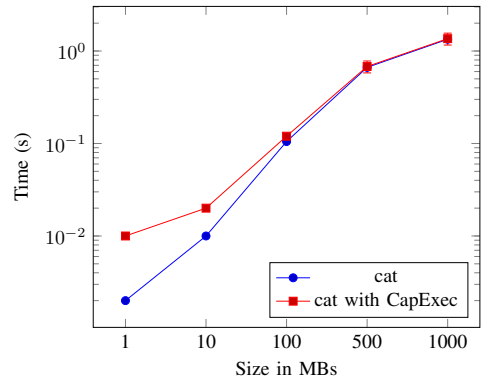

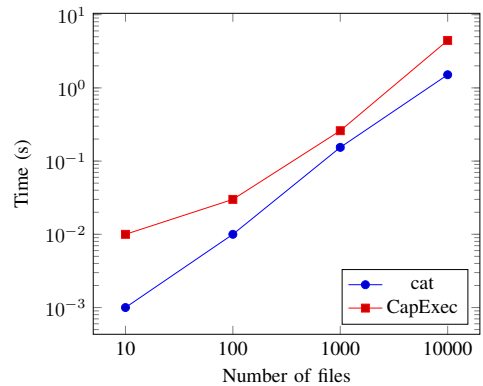
\begin{figure}[ht]
\begin{tikzpicture}[xscale=0.75, yscale=0.75]
\begin{axis}[
	ymin=0,
	ylabel={Time (s)},
	ymode = log,
	xlabel={Number of files},
	xtick={1, 2, 3, 4},
	xticklabels={10, 100, 1000, 10000},
	legend entries={cat, CapExec},
	legend pos= south east
]
\addplot+[error bars/.cd,y dir=both,y explicit]
coordinates {
	(1, 0.001)  +- (0, 0)
	(2, 0.01)  +- (0,0)
	(3, 0.154)	+- (0,0.01)
	(4, 1.512)      +- (0,0.0)
};
\addplot+[error bars/.cd,y dir=both,y explicit]
coordinates {
	(1, 0.01) +- (0, 0)
	(2, 0.03) +- (0, 0)
	(3, 0.26) +- (0, 0)
	(4, 4.448) +- (0, 0.05)
};
\end{axis}
\end{tikzpicture}
\caption{Time to open multiple files with \name{cat(1)}}
\label{cat-multiple-files}
\end{figure}

To investigate running \name{cat(1)} with {\APP}, which only needs one Casper service (\name{system.fileargs}), we examined \name{cat(1)} and its sandboxed version with two test scenarios. 
In the first scenario, we examined invocations of \name{cat} with files of various sizes from 1 MB to 1 GB, shown in figure \ref{cat-single-file}.
Since a large portion of {\APP} overhead is spent pre-opening and holding handles to file descriptors, in the second test scenario, we ran our tests with a varied number of \emph{empty} files as the input set, from 10 to 10000, shown in figure \ref{cat-multiple-files}. All measurements were performed ten times per test case.

As can be seen in figure \ref{cat-multiple-files}, executing \name{cat} in our sandbox adds overhead to the execution time. 
We find this delay more tolerable as the number of inputs grows. 
The majority of the cost of using {\APP} is spent in setting up the sandbox.
This includes the reading of configuration files, the creating of Casper services and \name{fork(2)} calls to load and open them.  

Here is how we explain various points affected execution time with the observed latency. {\APP} parses a service declaration file and then defines essential Casper services with their specified limits.
Opening Casper services is always associated with calling \name{fork(2)} for each service.
After all service definitions, \name{CapExec} forks and executes the binary, in the child process, in capability mode. 
Hence, the cost of establishing a sandbox environment and Casper channels always exist, but just once during runtime, which is negligible for large inputs, in terms of size. It shows 1\% on average in our cases for huge files, and 1\% to 2\% for small cases.
However, this step costs more for a large number of files with equal sizes in our second test scenario. In worst cases of those tests, it takes over twice of the time. We can explain this issue with this fact that there is a linear relationship between the number of mentioned services and their features in service declaration file and the time required for configuration.

There are additional reasons for the delays seen in our outputs.
The presence of preloaded libraries redirected system calls and the communication between the sandboxed process and Casper daemon, all affect process runtime. The increasing number of Casper services opened by the supervisor program causes more latency during runtime.

Although we believe these delays are acceptable to achieve secure services, there are various places in {\APP} which can be optimized.  For example, with better sets of service declaration options, such as Casper \name{fileargs} arguments, we could speed the first phase of sandbox creation. {\APP} has been developed with the most straightforward possible design as a proof of concept but additional optimization on the current system is one of our current focuses.

\begin{figure}[ht]
\begin{tikzpicture}[xscale=0.75, yscale=0.75]
\begin{axis}[
	ymin=0,
	ylabel={Memory (Mb)},
	%ymode = log,
	xlabel={Number of files},
	xtick={1, 2, 3, 4},
	xticklabels={10, 100, 1000, 10000},
	legend entries={cat, CapExec},
	legend pos= north west
]
\addplot+[error bars/.cd,y dir=both,y explicit]
coordinates {
	(1, 2.1148)  +- (0, 0.05)
	(2, 2.1327)  +- (0,02)
	(3, 2.1597)  +- (0,0.004)
	(4, 2.3551)  +- (0,0.008)
};
\addplot+[error bars/.cd,y dir=both,y explicit]
coordinates {
	(1, 4.2118) +- (0, 2.5834)
	(2, 5.5783) +- (0, 2.5327)
	(3, 7.8072) +- (0, 1.548)
	(4, 12.0822) +- (0, 1.8636)
};
\end{axis}
\end{tikzpicture}
\caption{Memory utilization for various number of files opened by \name{cat(1)}}
\label{cat-multiple-files-memory}
\end{figure}
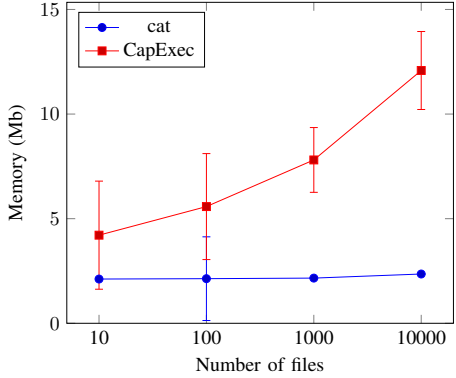

{\APP}'s approach includes the additional cost of spawning Casper services and {\APP}'s preloaded libraries. In addition to the structures keeping services configurations and limits, new processes have to be spawned by the supervisor program. We can see the impact of this design in figure \ref{cat-multiple-files-memory}, which shows a broader impact on memory overhead.
\\
\subsubsection{Completeness and Correctness}

We have investigated the correctness of our sandboxed application through \name{ktrace(1)} output, which provides us detailed information about invoked system calls and their returned values. For example, in the listing \ref{ktracecat} which is presenting the \name{ktrace(8)} output of the unmodified version of \name{cat(1)} program,  we can see the disallowed system call, \name{open(2)}, which is forbidden in the capability mode. 

\begin{lstlisting}[language=csh, caption={The trace of the syscall for the cat(1) application.}, label={ktracecat}]
cat    CALL  openat(AT_FDCWD,0x80024a010,0<O_RDONLY>)
cat    NAMI  "test/0"
cat    RET   openat 3
\end{lstlisting}

In Capsicum, when an application tries to access a global namespace, a capability error is returned, and the intended system call fails. Listing \ref{ktracecapcat} is demonstrating the capability error returned from the system call \name{open(2)} in capability mode.

\begin{lstlisting}[language=csh, caption={The trace of the syscall for the sandboxed application.}, label={ktracecapcat}]
cat    CALL  cap_enter
cat    RET   cap_enter 0
cat    CALL  openat(AT_FDCWD,0x7fffffffe990,0<O_RDONLY>)
cat    CAP   restricted VFS lookup
cat    RET   openat -1 errno 94 Not permitted in capability mode
\end{lstlisting}

\name{CapExec} delegates tasks to Casper services for disallowed replaced system calls. Requests are sent to Casper services through a capability channel which passes commands through UNIX domain sockets to Casper services. 
We can see this procedure in the \name{ktrace(8)} output shown in the listing \ref{ktracecapexeccat}.
The system call \name{sendto(2)}, instead of the forbidden system call, is invoked. We can see that the final \name{open(1)} system call was executed by a different process.
Since in our examined applications, no system call failed and no related capability error was observed and they worked normally, we can infer that the corresponding configuration file sufficed for the applications.

\begin{lstlisting}[language=csh, caption={The trace of the syscall for cat(1) under the CapExec.}, label={ktracecapexeccat}]
cat	CALL  sendto(0x4,0x80163b0a0,0x4e,0,0,0)
cat	GIO   fd 4 wrote 78 bytes
       0x0000 6c00 0200 0000 0000 0000 003b 0000 0000 0000 0004 0400 0500  |l..........;............|
       0x0018 0000 0000 0000 0000 0000 0000 0000 636d 6400 6f70 656e 0004  |..............cmd.open..|
       0x0030 0500 0700 0000 0000 0000 0000 0000 0000 0000 6e61 6d65 0074  |..................name.t|
       0x0048 6573 742f 3000                                               |est/0.|

cat	RET   sendto 78/0x4e
[...]
capexec   CALL  openat(AT_FDCWD,0x800283078,0<O_RDONLY>)
capexec   NAMI  "test/0"
\end{lstlisting}

\subsubsection{Comparisons with Virtualization-based Solutions}

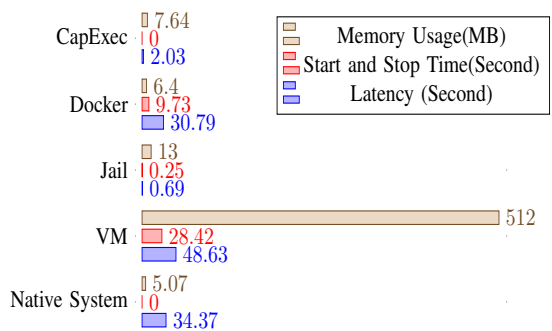
\begin{figure}
\begin{tikzpicture}[xscale=0.75, yscale=0.9]
  \begin{axis}[
    xbar,
    bar width=0.2cm,
    reverse legend,
    y axis line style = { opacity = 0 },
    axis x line       = none,
    tickwidth         = 0pt,
    enlarge y limits  = 0.2,
    enlarge x limits  = 0.02,
    symbolic y coords = {Native System,VM,Jail,Docker,CapExec},
    nodes near coords,
    legend style={at={(0.75,0.65)}, anchor=south,legend columns=1},
    width  = 0.45*\textwidth,
  ]
  
  \addplot coordinates { (2.026,CapExec)         (0.688,Jail)
                         (30.79,Docker)  (48.628,VM)  (34.37,Native System)};
   
  \addlegendentry{Latency (Second)}
  
  \addplot+[xbar] plot coordinates{ (0,CapExec)         (0.252,Jail)
                         (9.7324,Docker)  (28.422,VM) (0,Native System)};
    \addlegendentry{Start and Stop Time(Second)}
  
  \addplot coordinates { (7.64,CapExec)         (13,Jail)
                         (6.4,Docker)  (512,VM) (5.073,Native System)};
  \addlegendentry{Memory Usage(MB)}

  %\legend{Latency (Second), Memory (MB)}
  \end{axis}
\end{tikzpicture}
\caption{Comparison of running \name{traceroute} in different environments}
\label{traceroute-comparison}
\end{figure}

We have also compared {\APP} with equivalent virtualization-based solutions. There are various aspects to compare these systems such as required time to set up the sandboxed or virtualized environment, the complexity of configurations, required storage for each approach, memory utilization for each tool and finally the latency of running an application using each tool. To investigate \APP{} against most widely-used containers, we examined running \name{traceroute(8)} under five situations. First, we ran \name{traceroute(8)} on the native system out of any container or {\APP}. For the rest of the cases, we started the container, ran \name{traceroute(8)} on it and then stopped it. We practiced this procedure on a virtual machine, a FreeBSD \name{jail}, a \name{docker}, and finally with {\APP}. Figure \ref{traceroute-comparison} shows the latency and memory used of running \name{traceroute(8)} using each technology, giving an average of five test runs. It also shows the time spent to start each container up and to stop them after the test.
Since {\APP} runs directly on the native operating system, we can see that the memory overhead is negligible. 

There are also additional storage requirements for the investigated isolation techniques. As an example, the size of the image for a virtual machine, a docker image, and a jail image in our tests, were 2.6 GB, 204 MB, and 800 MB respectively, which shows that how much isolating a small application might cost. This issue makes a light user-level sandboxing application like \APP{} distinguished from other sandboxing solutions.

Regardless of numerical comparisons, we can compare the mentioned solutions in terms of some other non-statistical criteria. Here are some of the most significant aspects. 
Setting a virtual machine, a \name{jail(8)} or a \name{docker(8)} container require considerable time. 
The time to learn, configure, and run the first instance of these containers, depends on the underlying operating system and the user's proficiency.
For the various secured environment, one should define, create, and install various images to be virtualized later. 
In most cases, there are also post configurations to make the environment efficient to run the intended application. For example, to run network services under any of these applications, one might do many networking and firewall configurations to make the service and its response reachable through network. 
This issue holds for \APP{} as well but in a tremendously lightened way. We believe the time for setting up a sandbox and running a user-level application, is much less than creating and configuring each of those containers. 
To run an application under \APP{}, one should first find all disallowed system call invoked by it, and this stage could be complicated for unprofessional users. However, we tried to ease this step with \name{CapCheck}. 

As the last significant point, all of the described solutions above need the \name{root} privilege to run. While we are trying to create and run a sandbox under an unprivileged application. On the other hand, they are still probable to get compromised.  As an example, a virtual machine is itself breakable, so in that case, the network service running in it will be out of access. To avoid breaches in the future, a solution is to combine services running in virtualized containers with other services running out of them; however, this will add more complexity and effort to the problem.

\section{Related Work}
\label{sec:related-work}
To design \APP{}'s scheme, we have studied several sandboxing mechanisms and service managers, focusing on their security options. 
As the most important options, user permissions and privilege are essential   for most of contemporary \name{init} systems. 
Configuring \name{uid} and \name{gid} in \name{systemd}\cite{systemd}, \name{s6}\cite{s6}  or other service managers, are examples of these options.
However, they are still susceptible to complex attacks such as those utilizing privilege escalation.

\name{launchd}\cite{launchd}, the service manager on macOS, was the first system that expanded \name{inetd}'s\cite{inetd} socket activation, and was adopted because of its performance.
\name{launchd} provides options for security that are mostly focused on permission features such as \name{username}, \name{groupname}, \name{initgroup} and \name{umask}.
The only option concerning sandboxing schemes is the ability to set the \name{root} directory. The short list of security options in \name{launchd} originated from the internal security scheme of macOS, which is consistent and well-designed.

\name{systemd}\cite{systemd} is another widely-used service manager in which various security options are supported. In contrast with \name{launchd},
\name{systemd} supports very fine-grained security controls such as \name{uid}/\name{gid} control and isolation options such as inaccessible or read-only paths, root, and \name{tmp} directories.
Benefiting from \name{seccomp}\cite{seccompbpf}, system call filtering is also provided.
However, with all of these features, it is left to the developer to verify the configuration's compatibility with the use of the application but it is easy to define inappropriate policies.

There are also other service managers with similar security mechanisms such as \name{relaunchd}, also known as \name{jobd}\cite{jobd}, \name{nosh}\cite{nosh}, and \name{s6}\cite{s6}, which are quite different in design. 
For example, \name{relaunchd} runs services in jails \cite{jail}, and its options are configurable for the user, while most of \name{nosh}'s security features are internal. Also, \name{nosh} uses the concept of capabilities in design. \name{nosh}’s design and mechanism is based on \name{daemontools} which is a package of tools for UNIX service management\cite{daemontools-family}. 
There are also other service management tools inspired by \name{deamontool} such as \name{runit}\cite{runit} and \name{s6} that are much more than an init system. \name{s6} is a package of tools including various security features such as access control management on client connections, supporting \name{uid}-less privileges, ability to define \name{sudo} family \cite{s6-overview}, etc. Most of \name{deamontool}-based service managers benefit from internal security in their design.

In addition to service managers, we also have studied widely-used existing sandboxing tools such as  \name{chroot(2)} as a sandboxing system call, FreeBSD's \name{jail} and \name{docker} that were examined as containers in section \ref{evaluation}, \name{seccomp(2)}\cite{seccompbpf} as a framework that applies system call filtering, \name{CloudABI}\cite{cloudabi} which provides process-level sandboxing benefiting from capabilities, etc. An interesting point about all of these applications is the different level at which each  of them are providing sandboxing.

\section{Future Work}
\label{sec:future-work}

\APP{} is a service supervisor that executes one application at a time.
That application executes in isolation, a sandbox provided by Capsicum with limited access to global resources mediated by Casper.
However, \APP{} is not a complete service manager.
Our intent is to use \APP{} as a foundation for a service manager
that handles interdependencies and sandboxes groups of network services.
To aid in this goal, we are also developing more network-oriented Casper services.
Additionally, we are investigating ways to reduce the amount of security specific knowledge required to use {\APP}. We aim to make defining security policy as simple as specifying white-listed resources required by services, so that \APP{} decides which Casper services, with what configuration, should start. 

\section{Conclusion}

In this paper, we introduced \APP{}, a prototype sandboxing supervisor that facilitates service sandboxing both on local and network services.
Using Capsicum and Casper, along with a simple configuration file, we transparently provide this isolation at run time without any modification on the application's source code.
The system requires to be configured based on essential Casper services, which is challenging for the user, but to facilitate this procedure, we provide \name{CapCheck}, a tool to discover system calls that require wrapping.
This demonstrates that sandboxing itself can be a service, a key foundation for building security-aware service managers in the future.

\bibliographystyle{bib/IEEEbibtran}
\bibliography{bib/mybib}

\end{document}